\documentclass[letter]{aa} 
\usepackage{graphicx}
\usepackage{txfonts}
\usepackage[]{natbib}
\bibpunct{(}{)}{;}{a}{}{,}
\begin{document}

   \title{The obscured hyper-energetic GRB\,120624B \\hosted by a luminous compact galaxy at \textit{z}=2.20\thanks{Based on observations collected at the European Southern Observatory, Chile, with programmes 089.D-0256 and 090.D-0667, at the Gran Telescopio Canarias, with programmes GTC49-12A and GTC58-12B, at the Submillimeter Array with programme 2012A-S001, at CAHA with programme F13-3.5-031, at Liverpool Telescope with programme CL13A03 and a DDT programme at the Chandra X-ray Observatory.}}


   \author{A.~de~Ugarte~Postigo
          \inst{1,2}
          \and
          S.~Campana\inst{3}
          \and
          C.C.~Th\"one \inst{1}
          \and
          P.~D'Avanzo \inst{3}
          \and
          R.~S\'anchez-Ram\'irez\inst{1}
          \and
          A.~Melandri \inst{3}
	\and
	J.~Gorosabel \inst{1,4,5}
	\and
	G.~Ghirlanda \inst{3}
	\and 
	P.~Veres \inst{6}
	\and
	S.~Mart\'in \inst{7}
	\and
	G.~Petitpas\inst{8}
	\and
	S.~Covino \inst{3}
	\and
	J.P.U.~Fynbo \inst{2}
	\and
	A.J.~Levan \inst{9}
          }

   \institute{Instituto de Astrof\' isica de Andaluc\' ia (IAA-CSIC), Glorieta de la Astronom\' ia s/n, E-18008, Granada, Spain.\\
                     \email{adeugartepostigo@gmail.com}
         \and
             Dark Cosmology Centre, Niels Bohr Institute, Juliane Maries Vej 30, Copenhagen \O, D-2100, Denmark.
         \and
             INAF - Osservatorio Astronomico di Brera, via E. Bianchi 46, 23807, Merate, LC, Italy.
         \and
         	    Unidad Asociada Grupo Ciencia Planetarias UPV/EHU-IAA/CSIC, Departamento de F\'{\i}sica Aplicada I, E.T.S. Ingenier\'{\i}a, Universidad del Pa\'{\i}s Vasco UPV/EHU, Alameda de Urquijo s/n, E-48013 Bilbao, Spain.
	\and
	    Ikerbasque, Basque Foundation for Science, Alameda de Urquijo 36-5,E-48008 Bilbao, Spain.
	\and
	    Department of Astronomy and Astrophysics, Department of Physics, and Center for Particle and Gravitational Astrophysics,
Pennsylvania State University, 525 Davey Lab, University Park, PA 16802, USA.
         \and
             European Southern Observatory, Vitacura Casilla 19001, Santiago de Chile 19, Chile.
         \and
	    Harvard-Smithsonian Center for Astrophysics, Submillimeter Array, 645 North A'ohoku Place, Hilo, HI 96720, USA.
	\and
	    Department of Physics, University of Warwick, Coventry, CV4 7AL, UK.
	}
   \date{Received ; accepted }
 
  \abstract
   {Gamma-ray bursts are the most luminous explosions that we can witness in the Universe. Studying the most extreme cases of these phenomena allows us to constrain the limits for the progenitor models.}
   {In this Letter, we study the prompt emission, afterglow, and host galaxy of GRB\,120624B, one of the brightest GRBs detected by \textit{Fermi}, to derive the energetics of the event and characterise the host galaxy in which it was produced.}
   {Following the high-energy detection we conducted a multi-wavelength follow-up campaign, including near-infrared imaging from HAWKI/VLT, optical from OSIRIS/GTC, X-ray observations from the Chandra X-ray Observatory and at sub-millimetre/millimetre wavelengths from SMA. Optical/nIR spectroscopy was performed with X-shooter/VLT. }
   {We detect the X-ray and nIR afterglow of the burst and determine a redshift of $z=2.1974\pm0.0002$ through the identification of emission lines of [\ion{O}{ii}], [\ion{O}{iii}] and H-$\alpha$ from the host galaxy of the GRB. This implies an energy release of E$_{iso,\gamma}=(3.0\pm0.2)\times10^{54}$ erg, amongst the most luminous ever detected. The observations of the afterglow indicate high obscuration with $A_V>1.5$. The host galaxy is compact, with R$_{1/2}<$1.6 kpc, but luminous, at L$\sim$1.5 L$^*$ and has a star formation rate of $91\pm6$ M$_\odot$/yr as derived from H$\alpha$.}
   {As other highly obscured GRBs, GRB\,120624B is hosted by a luminous galaxy, which we also proof to be compact, with a very intense star formation. It is one of the most luminous host galaxies associated with a GRB, showing that the host galaxies of long GRBs are not always blue dwarf galaxies, as previously thought.}

   \maketitle
%

\section{Introduction}

Long gamma-ray bursts (GRBs) are extreme phenomena, that release isotropic equivalent energies in the range of $10^{51}-10^{54}$ erg in tens of seconds. This energy is thought to be emitted through collimated jets, which help to moderate the emitted energy to the order of $10^{51}$ erg. Assuming this range of energies, GRBs are commonly explained by the collapse of very massive stars \citep{woo93}, although there are theories such as those where a highly magnetised and rapidly rotating neutron star could also produce GRBs \citep{uso92}. 
There are a few cases where extreme energy releases of E$_{iso}>10^{54}$ ergs are produced. It has been suggested that some of these events have corrected energy releases that greatly exceed $10^{51}$ erg, which compromise some progenitor models \citep{cen11}. 

Extremely luminous events normally have very luminous counterparts. Even in these cases, some GRBs present very highly extinguished afterglows, due to dust in the line of sight. This can result in strong observational biases in the samples that we are studying \citep{fyn09,cov13}. Recent analysis of the hosts of obscured bursts have revealed that these events are produced in galaxies that are much more massive and luminous than previously thought \citep{kru11,ros12,per13}. 

GRB\,120624B was one of the brightest GRBs detected by the \textit{Fermi} satellite. In spite of this, initial searches failed to detect an optical counterpart due to strong intrinsic obscuration. In this Letter we present observations of its afterglow and luminous host galaxy, aiming at understanding the energetics of the burst and the environment in which it was produced. Throughout this work we assume a cosmology with $\Omega_m=0.27$, $\Omega_\Lambda=0.73$, and $H_0=71$ km s$^{-1}$ Mpc$^{-1}$.


\section{Observations}

\subsection{Prompt $\gamma$-ray emission}

GRB\,120624B was detected by the Gamma-Ray Burst Monitor \citep[GBM,][]{mee09} onboard the \textit{Fermi} satellite at 22:23:54.92 UT on 24 June 2012 \citep[t$_0$ hereafter;][]{gru12}. It was an extremely bright $\gamma$-ray event with a fluence of $(1.916\pm0.002)\times10^{-4}$ erg cm$^{-2}$ \citep[10keV-10MeV; ][]{gru12b}, which was also detected by  \textit{INTEGRAL}/SPI-ACS \citep{gru12}, Konus/\textit{WIND} \citep{gol12}, \textit{Swift}/BAT \citep{sak12}, and \textit{Suzaku}/WAM \citep{sak12_}. This fluence places it amongst the top 0.6\% of bursts detected by \textit{Fermi}/GBM and the 1.5\% of the bursts detected by \textit{Swift}/BAT \citep{dep12}. The emission of the burst extended up to very high energies, with a significant detection over 100 MeV by the \textit{Fermi}/LAT detector \citep{via12}. The T$_{90}$ duration of the burst is $\sim271$ s, placing it on the high end of the GRB duration distribution.

Analysis of the GBM data\footnote{http://heasarc.gsfc.nasa.gov/W3Browse/fermi/fermigbrst.html} shows 3 bright pulses, the first one starting 270 s before t$_0$ and the last one ending 29 s after t$_0$. 
We analysed the spectrum integrated over the total duration ($\sim T_{90}$) of the burst extracting the spectrum of the three main peaks. The background spectrum was obtained by selecting two time intervals before and after the burst. The sequence spectra in these two selected intervals were fitted with a third order polynomial to account for the possible time variation of the background spectrum. 
Both time integrated and time resolved spectra were fitted with the typical Band function \citep{ban93}. The best fit parameters of the average spectrum and of the three spectra corresponding to the main peaks are reported in Table~\ref{table:1}. 
In the non-standard analysis of the LAT data (LAT low energy events, LLE, roughly from 10 to 200 MeV), we also find the three peaks. In order to describe the high energy spectrum, the GBM and LAT measurements were combined. This shows that the spectrum in the LAT energy range is still well described by a Band function and no emerging high-energy extra component is evident. After the repoint, there was significant emission detected by standard analysis as well. We find the source with decreased flux density at 100 MeV compared to the emission seen in LLE data.  While the number of photons is low, the temporal evolution of the high energy portion appears broadly consistent with a power law decay.

\subsection{Afterglow observations}

The first optical observations that were reported, starting 22 hr after the burst were affected by the nearby Moon, and did not reach very constraining limits \citep{san12,xu12}. The deepest optical limits were obtained with the GROND multi-band imager \citep{gre08} at the 2.2m telescope at La Silla Observatory, one day after the burst, with \textit{r'}$<24.0$ \citep{sch12}, showing that the afterglow was already very faint and/or significantly extinguished.

Our observations began 24.7 hr after the burst using the near infrared (nIR) imager HAWKI at the Very Large Telescope (VLT), at Paranal Observatory. We performed observations using $K_S$-band, which were repeated the day after to check for variability, and again on day 19. Optical observations were done with OSIRIS at the GTC telescope 4.8 d after the burst in the $r$ band. The results of the photometry of these observations are shown in Table~\ref{table:obslog}.

The \textit{Swift}/XRT observations 2.5 d after the burst led to the identification of two faint sources within the BAT error circle \citep[S1 and S2; ][]{sak12}. Subsequent visits at 3.3 and 8.2 d detected only S2 but failed to impose strong-enough limits to confirm the decay of S2.

\textit{Chandra} observed the field of GRB\,120624B starting on June 30, 2012 at 23:58 UT, 6.1 d after the burst trigger. The field was imaged with the S3 CCD for 10 ks. A comparison with a \textit{Swift/}XRT image lead to the identification of the afterglow \citep{cam12}: S1 dimmed considerably and instead source S2 was at a comparable level. 
The position determined with {\tt wavdetect} is RA(J2000): 11$^{\rm h}$ 23$^{\rm m}$ 32$^{\rm s}$.32 and Dec(J2000): +8 $55^{\prime}$ $42\farcs8$. The uncertainty is dominated by the \textit{Chandra} boresight and is $0\farcs5$. 
The number of photons collected with \textit{Swift/}XRT and Chandra is very low. For this reason we used C-statistics within XSPEC. We fit the data assuming the same spectral model and leave free only the normalisation. We used an absorbed power law with a Galactic absorption component (fixed to $4.1\times 10^{20}$ cm$^{-2}$) and a free component at the host redshift. The best fit photon index is $\Gamma=2.7^{+1.2}_{-0.9}$ (obtained with $\Delta C=2.71$) and the intrinsic column density is $N_H(z)=2.7^{+4.6}_{-2.6}\times 10^{22}$ cm$^{-2}$ at the redshift of the GRB (see below). 

In view of the extreme brightness of the GRB event, we performed two observations in mm/submm wavelengths using the SubMillimeter Array (SMA).
The first observation, was carried out on June 26th between UT 4:30 and 8:30.
Observations were aimed at the early \textit{Swift}/BAT position, which was $18^{\prime\prime}$ away from the nominal GRB position, right in the edge of the $36^{\prime\prime}$ primary beam of the SMA antennas at the 358~GHz observed frequency.
The achieved 1$\sigma$ rms of 3.8~mJy yielded an r.m.s. of 11~mJy at the GRB position.
The second observation, was carried out on June 29th between UT 4:45 and 8:10, towards the position
RA(J2000): 11$^{\rm h}$ 23$^{\rm m}$ 31$^{\rm s}$.8 and 
Dec(J2000): +8 $55^{\prime}$ $47$ which showed tentative emission in the first observation.
This position resulted to be only $8^{\prime\prime}$ away from the GRB position, and well within the $55''$ primary beam of the SMA at the 230~GHz frequency of this second observation.
The observation resulted in a $1~\sigma$ r.m.s. of 0.7 mJy.
The limits obtained in our data are consistent with the 1mJy detection obtained at Plateau de Bure
Observatory in the 86GHz band \citep{bre12}.

\subsection{Host galaxy observations}

Late imaging was performed several months after the GRB using OSIRIS at the 10.4 m GTC and the 2.0 m Liverpool Telescope for the optical, HAWKI at the 8.2 m VLT for the nIR and Omega2000 at the 3.5 m CAHA in $Y$-band to study the host galaxy of the event, characterising its spectral energy distribution. The host is detected in the $g$, $r$, $i$, $z$, $J$, $H$ and $K_S$ filters as a point-like source, even in the highest resolution nIR data, where the seeing was $\sim0.4^{\prime\prime}$. The detections in the nIR have low signal-to-noise ratio (S/N), so an extended, lower surface brightness component could not be discounted, although no extended emission is either seen in the optical data, where the S/N is higher, with a seeing of $\sim1.1^{\prime\prime}$.

\begin{figure}[h!]
   \centering
   \includegraphics[width=7cm]{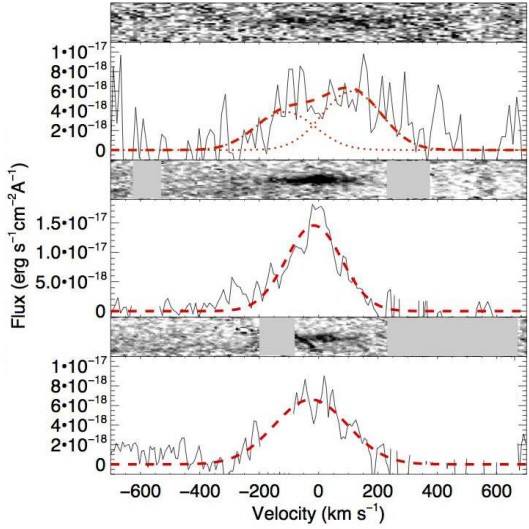}
      \caption{Sections of the X-shooter spectrum showing the emission lines of [\ion{O}{ii}] (top, centred between the two lines of the doublet), [\ion{O}{iii}] (middle) and \ion{H}{-$\alpha$} (bottom). Regions affected by sky lines and detector defects have been masked.
              }
         \label{Fig:em}
   \end{figure}

Between 17.1 and 19.1 days after the burst we performed 3 spectroscopic observations using the X-shooter spectrograph \citep{ver11} at the 8.2 m UT 2 of the VLT covering the complete wavelength range from 3000 to 24800 {\AA} at intermediate resolution. 
The total observation time consisted of $5\times$1200s in the UltraViolet/Blue and the Visible arms and $21\times$300 s in the NIR arm, with 1.0$^{\prime\prime}$, 0.9$^{\prime\prime}$ and 0.9$^{\prime\prime}$ slits, respectively. All the 3 epochs were combined together in order to produce the best possible spectrum, given that the observations had to be performed at high airmass and that the target was faint. The resulting spectrum has a seeing of 0.9$^{\prime\prime}$. In the combined data we detect emission lines of [\ion{O}{ii}], [\ion{O}{iii}] and \ion{H}{-$\alpha$} at a common redshift of $z=2.1974\pm0.0002$, which we identify as the redshift of the host galaxy and, consequently of the GRB (see Fig.~\ref{Fig:em}).

\section{Results and discussion}

\subsection{Afterglow}

Although image subtraction of the first two nIR epochs did not show any significant variability at the location of the GRB \citep{dav12, deu12b}, the comparison with the later epochs reveals the presence of a counterpart. Image subtraction was performed using the ISIS package \citep{ala98,ala00} with the $K_S$-band frames obtained with HAWKI. 
The difference in flux observed for our target between the first and the last epoch (which has only flux from the host galaxy) enabled us to estimate the pure afterglow contribution. 
In the first epoch, 1 d after the burst, the afterglow magnitude was $21.10\pm0.33$ and 2 d after the burst $21.40\pm0.40$, whereas 19 days after the burst, the afterglow is no longer detected. 
The location of the afterglow has an offset with respect to the centroid of the host galaxy of $0.11\pm0.04^{\prime\prime}$, equivalent to $0.94\pm0.34$ kpc at the redshift of the GRB. This is also consistent with the median offset of $\sim1.2$ kpc measured between long GRB afterglows and their host galaxies \citep{blo02,deu12c}. Our best equatorial coordinates of the afterglow, as compared to the SDSS DR8 catalogue, are (J2000 $\pm0.3^{\prime\prime}$, including statistical and catalogue errors) R.A.: 11$^{\rm h}$ 23$^{\rm m}$ 32$^{\rm s}$.30 and Dec.: +8 $55^{\prime}$ $42\farcs3$.

Considering our $K_S$-band and X-ray detections of the afterglow, we can make an estimate of the minimum extinction necessary to explain the detection limits provided by GROND \citep{sch12}. By assuming the least constraining case, where there is no spectral break between nIR and X-rays we get a maximum slope value to which we can add the Galactic extinction. We correct the detection limits of GROND by subtracting the contribution of the host galaxy, for which we have late time photometry. Under these conditions we find that we need an extinction of at least A$_V>1.5$ mag, to be consistent with the detection limits (see Fig.~\ref{Fig:sed}), assuming a Small Magellanic Cloud extinction law (the value is slightly higher if we consider a Milky Way extinction law). This confirms that GRB\,120624B is a strongly extinguished burst.
In this exercise we have combined data from slightly different epochs: HAWKI data correspond to day 2, GROND observations from day 1 and X-ray data are from 2.7 days after the burst. Our HAWKI observations show that there is not much difference between day 1 and day 2 in the nIR afterglow (the decay is $0.30\pm0.52$
 mag), extrapolating the GROND data to day 2 would just make the limits tighter. In a similar way, XRT is probably a bit lower than what would correspond, which would flatten the slope and again require stronger extinction. In conclusion, the A$_V>1.5$ mag is a conservative limit on the extinction of this burst.

With a peak emission of $\sim1$ mJy \citep{bre12}, the millimetre emission of GRB\,120624B is close to the typical luminosity in the sample of GRB afterglow detections \citep[$\sim10^{32}$erg s$^{-1}$ Hz$^{-1}$][]{deu12} and far from the extremely bright GRB afterglows, peaking at $\sim10^{33}$erg s$^{-1}$ Hz$^{-1}$.

\subsection{Host galaxy}

\begin{figure}[h!]
   \centering
   \includegraphics[width=\hsize]{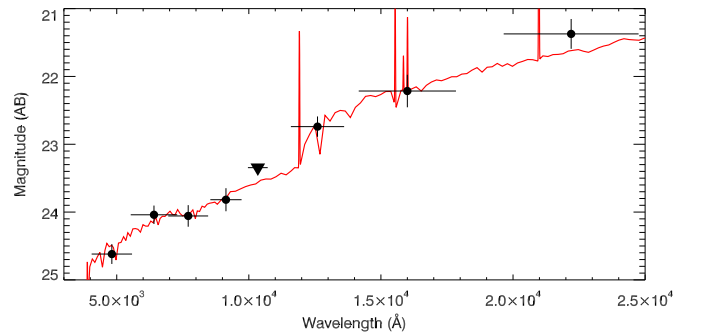}
      \caption{Optical to nIR flux density distribution of the host galaxy. Dots indicate the detections in $grizJHK_S$, whereas the triangle is a 3-$\sigma$ detection limit in $Y$. In red is the best fit to a galaxy template.
              }
         \label{Fig:host}
   \end{figure}

We use LePhare \citep[v. 2.2, ][]{arn99,ilb06} to fit the optical to NIR SED of the host to a set of galaxy templates based on the models from \citet{bru03} created following the same procedure as in \citet{kru11}. The SED shape is best reproduced ($\chi^2/d.o.f.=2.8/7$) by a galaxy template (see Fig.~\ref{Fig:host}) with an extinction of $E(B-V)=0.25$ \citep[Calzetti extinction law, implying $A_{V,host}=1.0$][]{cal00}, an age of $(1.0_{-0.8}^{+1.4})\times10^9$ yr, a mass of log(M$_*$/M$_\odot$)$=(10.6_{-0.2}^{+0.3})$ and a star formation rate of log(\textit{SFR})$ = (2.1_{-0.4}^{+0.5})$ M$_\odot$ yr$^{-1}$ (log(\textit{SSFR})$= (-8.4_{-0.7}^{+0.7})$ yr$^{-1}$). 

The host galaxy has an absolute magnitude of M$_r$=--22.2 mag and M$_g$=--21.6 (M$_r$=--23.1 and M$_g$=--22.6 extinction corrected). Taking the UV luminosity function at \textit{z}$\sim$2 \citep[e.g. ][]{red08} which corresponds to wavelengths between \textit{g} and \textit{r} in our observer frame, the galaxy has a luminosity of 1.5 L* assuming M$_{1700\AA}$=--22.0 mag, making it one of the most luminous GRB hosts detected to date \citep{kru11}. Our images show a very compact object consistent with being point-like, which at a seeing of 0$\farcs$4 in the best images implies a R$_{1/2}<$1.6 kpc. However, the emission lines are extended in the wavelength direction with a FWHM $\sim$200 km\,s$^{-1}$ following a smooth Gaussian profile and no extension in the spatial direction.

From the H$\alpha$ emission line (see appendix) we derive a SFR of $42\pm3$ M$_\odot$/yr ($91\pm6$ M$_\odot$/yr when corrected for extinction).
The errors do not include the uncertainties of the method \citep[$\sim30\%$][]{ken98} or the ones from the flux calibration ($\sim20\%$). 
We can also make a rough estimate on the metallicity using the R$_{23}$ parameter \citep{Kobulnicky} by extrapolating the H$\beta$ flux from the extinction corrected H$\alpha$ flux assuming case B recombination which demands a ratio of H$\alpha$/H$\beta$=2.86. This yields log R$_{23} =$ 0.84 which implies a metallicity of around 12+log(O/H)=8.4 or half solar (at this metallicity, the two-valued R$_{23}$ metallicity function has an unambiguous solution).

While most GRB hosts are low-mass blue galaxies \citep[e.g. ][]{Savaglio09}, recently a few massive hosts with luminosities and masses similar to that of the GRB\,120624B host have been found \citep{ros12, kru11, per13}. In fact, dark GRBs are more often found in massive galaxies, some of which are EROs \citep{ros12}, and often the extinction in the host galaxy correlates with the extinction of the afterglow, implying a rather uniform dust distribution in the host \citep{per13}. Most of those massive dusty GRB hosts also have supersolar metallicities (e.g. GRB\,020819, \citealt{Levesque10}; GRB\,080605, \citealt{KruehlerGRB}; GRB\,070802, \citealt{Eliasdottir}), naturally following the mass-metallicity relation. The host of GRB\,120624B, however, has a somewhat lower metallicity as expected for its mass at \textit{z}$\sim$2 \citep{Erb06} which could be related to the high star-formation as SF galaxies usually have lower metallicities \citep{Mannucci}. The afterglow has a somewhat higher extinction than the host, but still within the equal distribution for GRB hosts as shown in \citet{per13}, Fig. 15, although we have to consider that the extinction of the afterglow is a lower limit. GRB 120624B might therefore be in an intrinsically more extinct region like a molecular cloud that has not yet resolved or simply behind a more obscured sightline in the host.

The host of GRB\,120624B could be similar to luminous compact galaxies (LCGs) detected at somewhat lower redshifts and proposed to be the progenitors of today's spiral galaxy bulges \citep[e.g.][]{Hammer01}. LCGs have luminosities up to 1.4 L* and masses up to 10$^{11}$ M$_\odot$, half light radii of R$_{1/2}$<4.5\,kpc, large SFRs (average 40 M$_\odot$/yr), relatively high extinctions (A$_V$$\sim$1.5\, mag) and a range of metallicities all similar to our GRB host. Some LCGs show indications for a low surface brightness region surrounding the compact core and they all have strong metal absorption lines, both of which we would be unobservable in our case. Our host galaxy might hence be similar to a LCG which is in a crucial state of its evolution experiencing an episode of heavy SF before developing in a spiral or an elliptical galaxy.

\section{Conclusions}

   \begin{enumerate}
      \item We unambiguously identify the NIR and X-ray counterpart of GRB\,120624B. 
      \item Combining our afterglow detections with the available limits from the literature, we estimate an extinction larger than A$_V>1.5$ magnitudes.
      \item Spectroscopy of the host galaxy reveals strong emission lines at a redshift of $z=2.1975\pm0.0002$ and FWHM of $200$ km s$^{-1}$.
      \item At this redshift, GRB\,120624B was an extremely luminous event, with an E$_{iso,\gamma}=(3.0\pm0.2)\times10^{54}$ erg.
      \item The host galaxy can be classified as a LCG, with a R$_{1/2}<$1.6 kpc, and a luminosity corresponding to $\sim1.5$L*.
      \item It is one of the GRB hosts with higher SFR, with \textbf{$91\pm6$} M$_\odot$/yr as derived from H$\alpha$.
      
   \end{enumerate}

\begin{acknowledgements}
We thank the anonymous referee for the careful and constructive revision of our manuscript.
We thank Thomas Kr\"uhler for his help with the SED fitting and Lise Christensen for fruitful discussions.
Based on observations made with the Gran Telescopio Canarias (GTC), at Roque de los Muchachos Observatory (La Palma).
The research activity of AdUP, CT and JG is supported by Spanish research project AYA2012-39362-C02-02.
AdUP acknowledges support by the European Commission under the Marie Curie Career Integration Grant programme (FP7-PEOPLE-2012-CIG 322307).
JG is supported by the Unidad Asociada IAA-CSIC\_ETSI-UPV/EHU and the Ikerbasque Foundation for Science.
PV thanks NASA NNX13AH50G and OTKA K077795.
The Dark Cosmology Centre is funded by the DNRF.
SC acknowledges support by ASI grant I/011/07/0 and PRIN-MIUR grant 2009ERC3HT.
We thank H. Tananbaum for granting the Chandra DDT.

\end{acknowledgements}

\bibliographystyle{aa}
\bibliography{120624B}

  \Online

\begin{appendix}

\section{Prompt emission}

\begin{table*}[h]
\caption{GBM energy analysis of the 3 main $\gamma$-ray peaks and of the complete burst.}             
\label{table:1}      
\centering                          
\begin{tabular}{c c c c c c c}        
\hline\hline                 
t-t$_0$ interval 		&	$a_1$		&	$a_2$		& E$_{p,obs}$	& Fluence						& E$_{iso}$				& E$_{peak,rest}$ 	\\    
(s)				&				&				& (keV)		& (erg s$^{-1}$ cm$^{-2}$)		& (erg)					& (keV)			\\
\hline                        
(-270.34, -155.65)	& -1.1$\pm$0.04	& -2.35$\pm$0.30	& 516$\pm$73	& $(2.02\pm0.06)\times10^{-5}$	& $(9.2\pm0.03)\times10^{53}$	& $1592\pm320$	\\
(-126.98, -53.25)	& -0.93$\pm$-0.03	& -2.51$\pm$0.31	& 603$\pm$50	& $(1.06\pm0.02)\times10^{-4}$	& $(1.7\pm0.03)\times10^{54}$	& $1928\pm160$	\\
(-8.19, +28.67)		& -1.04$\pm$0.06	& -2.07$\pm$0.22	& 498$\pm$100&$(5.46\pm0.02)\times10^{-5}$	& $(8.5\pm0.3)\times10^{53}$	& $1650\pm233$	\\
\hline                                   
(-270.34, +28.67)	&				&			&				&							& $(3.0\pm0.2)\times10^{54}$	& $1809\pm64$	\\
\hline                                   
\end{tabular}
\end{table*}

Table~\ref{table:1} displays the results of the spectral fitting of the \textit{Fermi}/GBM data.

\section{Observations}

   \begin{figure}[h!]
   \centering
   \includegraphics[width=\hsize]{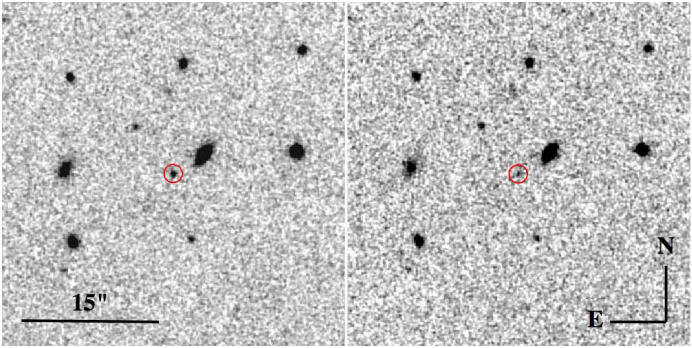}
      \caption{$K_S$-band images of the field of GRB\,120624B obtained 1 day after the burst (left), where the afterglow is still visible, and 6 months later (right) when only the host galaxy remains. The seeing in both frames is $\sim0.4^{\prime\prime}$.
              }
         \label{Fig:fc}
   \end{figure}

\begin{table*}
\caption{Observing log. Optical and nIR magnitudes and fluxes have not been corrected for Galactic extinction \citep[$E(B-V)=0.049$; ][]{sch11}. Optical data have been calibrated using SDSS as reference and for NIR we have used UKIDSS. }             
\label{table:obslog}      
\centering                          
\begin{tabular}{c c c c c c}        
\hline\hline                 
t-t$_0$ 	& Exp. 		& Instrument/ 	& Band 	& AB mag. 	& Flux density	\\    
(days)	& (s) 			& Telescope	&		&			&	(Jy)	\\
\hline                   
1.02		& 10$\times$60& HAWKI/VLT	& $K_S$	& $20.48\pm0.12$	& $(2.33\pm0.26)\times10^{-5}$	\\
2.05		& 10$\times$60& HAWKI/VLT	& $K_S$	& $20.64\pm0.13$	& $(2.01\pm0.24)\times10^{-5}$			\\
4.84		& 15$\times$15& OSIRIS/GTC	& $r$	& $>23.3$			& $<1.7\times10^{-6}$ \\
19.0		& 10$\times$60& HAWKI/VLT	& $K_S$	& $21.14\pm0.29$	& $(1.27\pm0.33)\times10^{-5}$			\\
\hline                   
277		& 5$\times$700 & LT	& $u$	& $>23.0$			& $<2.3\times10^{-6}$								\\
194		& 11$\times$200& OSIRIS/GTC& $g$	& $24.80\pm0.15$	& $(4.36\pm0.60)\times10^{-7}$			\\
194		& 16$\times$80& OSIRIS/GTC	& $r$	& $24.17\pm0.14$	& $(7.80\pm1.01)\times10^{-7}$			\\
194		& 6$\times$100& OSIRIS/GTC	& $i$		& $24.15\pm0.16$	& $(7.94\pm1.17)\times10^{-7}$			\\
194		& 18$\times$90& OSIRIS/GTC	& $z$	& $23.89\pm0.17$	& $(1.01\pm1.58)\times10^{-6}$			\\
252		& 93$\times$60 & OMEGA2000/3.5mCAHA	& $Y$	& $>23.4$			& $<1.6\times10^{-6}$								\\
203		&  20$\times$60& HAWKI/VLT	& $J$	& $22.78\pm0.15$	& $(2.81\pm0.39)\times10^{-6}$			\\
203		&  17$\times$60& HAWKI/VLT	& $H$	& $22.24\pm0.24$	& $(4.61\pm1.02)\times10^{-6}$			\\
203		&  17$\times$60& HAWKI/VLT	& $K_S$	& $21.39\pm0.22$	& $(1.01\pm0.20)\times10^{-5}$			\\
\hline\hline
1.34		& 9000		& SMA		& 358 GHz	& 			&	$<3.3\times10^{-2}$	\\
4.34		& 8172		& SMA		& 230 GHz	& 			&	$<2.1\times10^{-3}$	\\
\hline         
\end{tabular}
\end{table*}

Table~\ref{table:obslog} presents the observing log of our optical, nIR and millimetre/submillimetre observations.

\section{Spectral energy distribution and intrinsic extinction}

\begin{figure}[h!]
   \centering
   \includegraphics[width=\hsize]{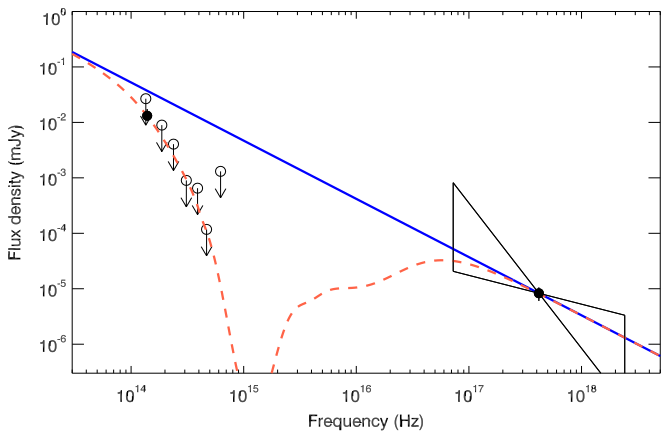}
      \caption{NIR to X-ray SED at $\sim$2 days, showing an estimation of the minimum extinction necessary to explain the detection limits. Filled circles indicate NIR and X-ray afterglow detections, whereas empty circles with arrows mark detection limits. The thick line is the minimum slope needed between NIR and X-rays to explain the detections and the dashed line is the same slope after adding the Galactic extinction and an intrinsic extinction of A$_V=1.5$ magnitudes.
              }
         \label{Fig:sed}
   \end{figure}

Figure~\ref{Fig:sed} shows the fit of the nIR to X-ray spectral energy distribution with an extinguished power law. From this fit we derive a minimum intrinsic extinction in the line of sight of the GRB of A$_V=1.5$ mag, as described in Section 3.1.

\section{Spectral line fluxes}

Table~\ref{table:lines} presents the results of fitting gaussians to the host galaxy emission lines. In the case of the $[\ion{O}{ii}] \lambda\lambda$3727, 3729 doublet and due to the low signal to noise ratio of the feature the width of the gaussian was fixed to $\sigma=100$ km s$^{-1}$ in order to constrain the number of free parameters in the fit. In all cases the spectral extractions were done with apertures of 1.4$^{\prime\prime}$, well above the 0.9$^{\prime\prime}$ seeing to minimise light losses. The flux errors do not include any systematic error in the absolute flux calibration which can be estimated to be of $\sim$20\%.

 \begin{table*}[h]
\caption{Fits to the emission lines of the host galaxy. The extinction correction is based on the photometric fit of the host galaxy, using a Calzetti extinction law with $E(B-V)=0.25$ mag, as described in the main text.}             
\label{table:lines}      
\centering                          
\begin{tabular}{c c c c}        
\hline\hline                 
Spectral line 					& Flux (erg s$^{-1}$ cm$^{-2}$)	& Corr. with $E(B-V)$			& $\sigma$ (km s$^{-1}$) 	\\    
\hline  
H-$\alpha \lambda$6564			& $(1.43\pm0.10)\times10^{-16}$			& \textbf{$(3.08\pm0.22)\times10^{-16}$}	&$123\pm11$\\
$[\ion{O}{iii}] \lambda$5008		& $(1.82\pm0.07)\times10^{-16}$			& \textbf{$(5.09\pm0.19)\times10^{-16}$}	& $93\pm4$ \\
$[\ion{O}{ii}] \lambda$3729		& $(6.0\pm0.3)\times10^{-17}$				& \textbf{$(2.31\pm0.12)\times10^{-16}$}	& $100$ (Fixed) \\
$[\ion{O}{ii}] \lambda$3727		& $(3.9\pm0.4)\times10^{-17}$				& \textbf{$(1.50\pm0.15)\times10^{-16}$}	& $100$ (Fixed) \\
\hline                                   
\end{tabular}
\end{table*}

\end{appendix}

\end{document}